\documentclass[prl,twocolumn]{revtex4-1}

\usepackage[dvips]{graphics,color}
\usepackage{graphicx}
\usepackage{subfigure}
\usepackage{amssymb}
\usepackage{color}
\usepackage{hyperref}
\usepackage[T1]{fontenc}
\usepackage[latin9]{inputenc}
\usepackage{amsmath}
\usepackage{setspace}

\usepackage{mathrsfs}

\def\be{\begin{equation}}
\def\ee{\end{equation}}
\def\e#1{\label{#1}\end{equation}}
\def\bea{\begin{eqnarray}}
\def\eea{\end{eqnarray}}
\def\ea#1{\label{#1}\end{eqnarray}}

\def\eqref#1{(\ref{#1})}
\def\bem#1{\begin{mathletters}\label{#1}}
\def\eml{\end{mathletters}}

\def\4#1{{\boldsymbol{#1}}}
\def\8#1{{\widetilde{#1}}}
\def\mean#1{{\langle{#1}\rangle}}

\begin{document}

\title{Work extraction via quantum nondemolition measurements of qubits in cavities: Non-Markovian effects.} 
\author{D. Gelbwaser-Klimovsky$^1$, N. Erez$^1$, R. Alicki$^{2}$ and G. Kurizki$^1$}
\address{
Department of Chemical Physics, Weizmann Institute of Science, Rehovot, 76100, Israel
\\
$^2$Institute of Theoretical Physics and Astrophysics, University of Gda\'nsk, Wita
Stwosza 57, PL 80-952 Gda\'nsk, Poland;  Weston Visiting Professor, Weizmann Institute of Science, Rehovot 76100, Israel\\
}

\begin{abstract}
We show that  frequent nondemolition measurements  of  a  quantum system immersed in a thermal  bath  allow the extraction of   work in a closed cycle from the system-bath \textit{interaction (correlation)} energy,  a hitherto unexploited  work resource.   It allows for work even if \textit{no information} is gathered or the bath is at zero temperature, provided the cycle is within the bath memory time.
 The predicted work resource may be the basis of  quantum engines  embedded in a   bath with long memory time, such as the electromagnetic bath of a high-\textit{Q} cavity coupled to two-level systems. 
\end{abstract}
\maketitle


\textbf{I. Introduction}

Information acquired by an observer  on the system can   be commuted into work in  a single-bath engine \cite{Szilard,landauer,caves,scullyPRL2001,uedanature,segal2010,MaruyamaRMP2009,Ruschhauptjphisb06,morebook}. To this end, the observer (``Maxwell's demon'')  must manipulate  the system    according to the results of  measurements performed on it. The measurements  have therefore  to be \textit{selective}, i.e.,  their outcomes must be read  and discriminated to determine the observer's course of action \cite{MaruyamaRMP2009,uedatheory}. The  balance of work and information is embodied by the   Szilard-Landauer (SL) principle \cite{Szilard,landauer} whereby \textit{work obtainable from a measurement must not exceed the energy cost of erasing its record from the observer's memory.}   
Notwithstanding  ongoing   efforts  to expand information-based thermodynamics  (IT) so as to include measurement-cost effects \cite{uedatheory,JacobsPRE2012,delrioNAT11} beyond the original SL balance (see Appendix), an important aspect has  been  little addressed thus far \cite{AllahverdyanPRE05}: How essential is the \textit{thermodynamic paradigm of system-bath separability}\cite{Alickiwork,Lindbladbook}  to the analysis of work extraction by measurements?

This question is here investigated in the context of non-Markovian quantum thermodynamics ``under observation'' \cite{nature,chapinnat08,njp,njp10,Gonzalo,Jahnke}. Namely,  we show  that frequent measurements can induce changes in  \textit{ system-bath correlations}, unaccounted for by the separability paradigm, and thereby change the   tradeoff of information and work.
  Consequently, \textit{selective} (read) measurements can extract \textit{more} work in a closed cycle than what the SL principle allows.  This effect requires the cycle to be completed  \textit{within a non-Markovian} (bath-memory) \textit{time-scale}.  

   We further show that   \textit{non-selective}  quantum nondemolition (QND) measurements of the system energy  (which we shall also denote as \textit{unread measurements}, i.e., measurements whose outcomes do not matter) enable the system to  do work in a cycle, although they  provide \textit{no information} on the system, nor do they change its state (by their definition as QND measurements)  and thus do not act as a ``demon.'' 
 Work is shown to be obtainable  within the bath memory time even at zero temperature ($T=0$), i.e., for the vacuum state of the bath. This finding is in stark contrast to the expectation that the extractable work at $T=0$ should vanish by the SL principle\cite{Szilard,landauer,caves,scullyPRL2001,uedanature,segal2010,MaruyamaRMP2009,Ruschhauptjphisb06} or its current  generalizations \cite{uedatheory, JacobsPRE2012} that are valid in the Markovian limit.

The predicted effects cannot  be ascribed to  quantum  \textit{coherence} in the system, which is the source of work in recently explored quantum heat engines (QHEs) \cite{Scully2011,*Scullysinglebath,*scullysinglebath2}, since coherence is absent from this scenario. 
Rather, these  effects  follow up on the anomalous temperature  effects of  frequent  QND measurements  on non-Markovian time scales \cite{nature,chapinnat08,njp,njp10,Gonzalo,Jahnke}. Here, we show that  these effects may well determine the  performance  of quantum engines   that operate on such  time scales. Clearly, one could instead consider energy pumping into the sytem by non-QND measurements, e.g., projections onto the \textit{x}-\textit{y} plane of the qubit Bloch sphere. Yet we are interested in \textit{minimal}  intrusion into the system that  would have \textit{no effect} on an  \textit{isolated} qubit (in the absence of a bath).

To this end, we analyze the following simple yet unconventional protocol: A brief unread QND measurement of the energy of a thermalized qubit   neither changes its state nor yields information, yet inevitably decorrelates it from its bath and therefore requires energy investment that is absent in SL considerations (Sec. II). 
A post-measurement cycle produced by sinusoidal modulation of the qubit frequency  is shown (both analytically and numerically) to yield work (despite the fact that  no information is obtained) from the unread measurement, provided the cycle is completed within the bath memory time (Sec. III). Post-measurement system-bath correlation change is shown to be the work resource. The analysis culminates in a  \textit{revised work-information relation} (Sec. IV). A  demonstration of consistency with the second law (Sec. V) is followed by a discussion of the  cost of multiple cycles (Sec. VI) and an analysis of feasible cavity-based  experimental scenarios (Sec. VII). The findings are summarized in Sec. VIII.

\textbf{II. System-bath decorrelation via QND measurements}

We  consider a QND measurement of the energy of a thermalized qubit by a quantum probe (P) consisting of two degenerate yet distinguishable states ($H_P=0$) (e.g., photon-polarization states). This situation is modeled by the Hamiltonian

\begin{equation}
H_{tot}=H_S+H_B+H_{SB}+H_{SP}(t)
\label{eq:totham}
\end{equation}
where S labels a two-level system (TLS) with energy states $|e\rangle$ and $|g\rangle$, B denotes a (bosonic or fermionic) bath,  $H_{SB}$ is the
  coupling of  S to a   bath operator $\hat{B}$  and $H_{SP}(t)$  is the S-P  interaction that effects the measurement. We choose the coupling Hamiltonians to have the  form:  $H_{SB}=\sigma_x \hat{B}$, where $\sigma_x$ does not commute with $H_S$ and neither does $\hat{B}$ with $H_B$, in order to describe S-B equilibration, i.e., transitions between the levels of S and B. 
	
	 Since  the measurement  is to have a QND  effect on S \cite{ScullyBOOK97}, in order for the density matrix of S, $\rho_S$, to
retain \textit{the same $\sigma_z$-diagonal form it had in equilibrium}, one should
choose  a projective
measurement in the $\sigma_z$- (qubit-energy) basis, i.e., $H_{SP}\propto\sigma_z$. 
The impulsive QND measurement of the  energy of S by P is well described by \cite{nature,njp,njp10,Gonzalo} 
$
e^{-i\int_0^{\tau_m} dt \frac{H_{SP}(t)}{\hbar}}=U_C,
$
where $U_C$ is the  Controlled-NOT (CNOT) operation 
  (with the  probe acting as the target qubit). Its  duration $\tau_m$ is assumed much shorter than all other time scales of evolution generated by $H_{tot}$.

The time-dependent system-probe coupling  aimed at a CNOT operation has the form ($|0\rangle$ and $|1\rangle$ being the probe states)

\begin{equation}
H_{SP}(t)=h(t)\left|e\right>\left<e\right| \left(
\left|0\right>\left<0\right| +
\left|1\right>\left<1\right|-
\left|0\right>\left<1\right|-
\left|1\right>\left<0\right|
\right)
\label{eq:hsd}
\end{equation}
 
\noindent where we may choose 
 
\begin{equation}
h(t)=\frac{\pi}{4\tau_m}\left(tanh^2\left(\frac{t-t_m}{\tau_m}\right)-1\right)
\label{eq:tempdepdet}
\end{equation}
 
\noindent as a smooth temporal profile of the system coupling to the probe qubits during the measurement at time $t_m$ which lasts over time  $\tau_m$.
\\
 The measurement outcomes are averaged over (for nonselective measurements-NSM), by tracing out the probe degrees of freedom. As long as the probe state is of the form 
 
 \begin{equation}
 \rho_P=I+d\sigma_z
 \end{equation}
 
 \noindent where $d$ is real, the measurement will not affect $\rho_S$, which is diagonal in the energy basis:

\begin{eqnarray}
\rho_S\longmapsto Tr_P\left(U_C \rho_S \otimes  \rho_P U^{\dagger}_C\right)= \nonumber\\
\left|e\right> \left<e\right| \rho_S \left|e\right> \left<e\right|+
\left|g\right> \left<g\right| \rho_S \left|g\right> \left<g\right|
\label{eq:measrhos}
\end{eqnarray}

\noindent i.e., the diagonal elements of $\rho_S$ are \textit{unchanged}, and the off-diagonal elements are erased. Since
the system is entangled with the bath, the effect of the measurement on $\rho_{S+B}$ is:

\begin{eqnarray}
\rho_{S+B}\longmapsto Tr_P \left(U_C \rho_{S+B} \otimes \rho_P U^{\dagger}_C\right)= \nonumber \\
\left|e\right> \left<e\right| \rho_{S+B} \left|e\right> \left<e\right|+
\left|g\right> \left<g\right| \rho_{S+B} \left|g\right> \left<g\right|\equiv \nonumber \\
B_{ee}\left|e\right> \left<e\right|+B_{gg}\left|g\right> \left<g\right|
\label{eq:measrhotot}
\end{eqnarray}

\noindent where $B_{ee(gg)}$ are bath states correlated to $|e\rangle$ and $|g\rangle$ respectively. Thus, the post-measurement  $\rho_{S+B}$ is block-diagonal in the energy states of the system. It can be shown\cite{nature,njp10} to be close to a product state of $\rho_S$ and $\rho_B$.
We assume at this stage that the state of P is \textit{unread} (averaged out) after this measurement. The entire measuring process can be summarized as 

\begin{equation}
\rho_{tot}=\rho_P \otimes \rho_{S+B} \rightarrow \rho_{S+P}\otimes\rho_B; \quad
Tr_P \rho_{tot} \rightarrow \rho_S\otimes \rho_B.
\label{eq:proc}
\end{equation}

  If B effects were treated classically, or S-B correlations were ignored, this measurement would have no effect at all, since it does not change the energy-diagonal state of S. Yet, because of the non-commutativity of $H_{SP}$ and $H_{SB},$ an impulsive NSM does change the S+B ``supersystem''. Such change is absent from Markovian treatments wherein the measurement is assumed \textit{slow enough} to warrant energy conservation of the supersystem\cite{alicki2004thermodynamics}, in contrast to the present  \textit{fast one that breaks this conservation}. 
This  crucial point is beyond   the system-bath separability paradigm\cite{Lindbladbook,alicki2004thermodynamics},  and stems from the fact that at equilibrium S is \textit{correlated} with B: they are in a   Gibbs state 
$
\rho_{Eq}=\frac{e^{-\beta(H_S+H_B+H_{SB})}}{Z}
\label{eq:rhoeq}
$,
 $\beta$ being the inverse temperature, and their \textit{mean correlation energy is negative} to ensure stable equilibrium: $\langle H_{SB}\rangle_{Eq}<0$ \cite{nature,njp,njp10}.
 The impulsive NSM changes the Gibbs state and its mean observables into their post-measured  counterparts\cite{njp}  

\begin{eqnarray}
\langle H_{SB} \rangle_{Eq}<0 \mapsto
\langle H_{SB} \rangle = \nonumber \\
\frac{1}{2}\langle H_{SB} \rangle_{Eq} +  \frac{1}{2} {\rm Tr}[\hat{B}  \sigma_z \sigma_x \sigma_z \rho_{Eq}]=0,
\label{eq:enepm}
\end{eqnarray}

\noindent where we have used the identity $\sigma_z \sigma_x \sigma_z = -\sigma_x$. 
Concurrently, $\langle H_{SP} \rangle_{Eq}=0 \mapsto \langle H_{SP} \rangle=\langle H_{SB} \rangle_{Eq}$, and $\langle H_P \rangle=0$ both before and after the measurement. Hence, the decrease in  the S+P correlation energy $\langle H_{SP} \rangle$ reflects an equal increase in the correlation energy  $\langle H_{SB} \rangle$. 

The foregoing expressions  have the following meaning:
Any NSM must increase the  mixedness and thus the entropy of $\rho_{S+B}$ \cite{petruccione}.
Since prior to the measurement S+B was in a  maximal-entropy (Gibbs) state among all states with the same mean energy $\langle H_{SB} \rangle$, the entropy increase implies an increase of $\mean{H_{SB}}$. Since the QND NSM changes $\rho_{S+B}$ into $\rho_S\otimes \rho_B$, \textit{yet leaves $\rho_{S}$ and $\rho_{B}$ unchanged} ,  it changes neither $\langle H_{S}\rangle $ nor $\langle H_{B}\rangle$. 
The  work invested in  performing the impulsive NSM is therefore

\begin{equation}
\Delta E_{meas}=-\Delta\langle H_{SP}\rangle=-\langle H_{SB} \rangle _{Eq}>0.
\label{eq:workin}
\end{equation}

 Equation \eqref{eq:workin}  reveals the discrepancy between the energy cost (work investment) required for a \textit{brief} QND measurement and that expected by  the SL principle:   the SL work investment  in the measurement is $W_{SL}^{meas}=0,$  since it  \textit{does not account} for system-bath correlations. As shown in what follows, Eq. \eqref{eq:workin} can be a useful work resource only on non-Markovian time-scales at \textit{any temperature }of the bath, including $T=0$. By contrast, the measurement cost in current Markovian treatments\cite{uedatheory,JacobsPRE2012} vanishes at $T=0$ (see Appendix).

\textbf{III Work extraction in a  non-Markovian cycle}
 As shown below, it is crucial to ensure  that the cycle duration $t_{cycle}$ satisfies $t_c \geq t_{cycle}$ , i.e.,   work extraction must occur on the non-Markovian  (memory) time-scale $t_c$ of S+B.
Furthermore, it should  be much longer than the duration  $\tau_m$ of the measurement that triggers the cycle (Sec. II).

If these conditions hold, the \textit{maximal} work extractable in a post-measurement  optimal cycle is (see App.)

\begin{equation}
(W^{ext}_{NSM})_{Max}=\Delta E_{meas}-T\Delta \mathscr{S}_{meas},\label{eq:main}
\end{equation}

\noindent where $\Delta E_{meas}$ is given by Eq. \eqref{eq:workin}, and $\Delta \mathscr{S}_{meas}$ is the NSM-induced \textit{entropy increase} of the supersystem S+B. This entropy increase reflects the destruction of S-B correlations (off-diagonal elements of $\rho_{SB}$) and the lack of information gain by the NSM.

 The bound on the maximal extractable work discussed above \textit{does not suffice} to demonstrate that work is indeed obtainable from S in a post-measurement  cycle: we must show that  the extracted  work $W^{ext}_{cycle}>0$ in a feasible cycle. 
This work may be extracted by a classical, coherent (zero-entropy) off-resonant piston (harmonic oscillator) that is dispersively  coupled to S via $\sigma_Z$, and modulates its energy levels \cite{alicki2004thermodynamics}, allowing S to undergo a closed cycle, after which $\rho_S(t_{cycle})=\rho_S(0)$. The piston coherent excitation-change then expresses the extracted work (see Discussion).

   The standard    expression for the  \textit{extractable} work (i.e., the negative of the \textit{invested} work)   over a cycle is \cite{Alickiwork, alicki2004thermodynamics}:

\begin{eqnarray}
W^{ext}_{cycle}=-\oint Tr\{\rho_S\dot{H}_S\}dt=-\oint s(t)\dot{\omega}(t)dt.
\label{eq:energypart}
\end{eqnarray}

\noindent Here $\omega(t)$ is the level-separation (frequency) of the piston-driven TLS, $s(t)$ is  the polarization (population difference) of  its energy-states  $\left|e\right>$, $\left|g\right>$ and   the cyclic integral is over a closed  trajectory in the frequency-polarization plane.

According to  \textit{the standard}  (Markovian) expression of the second law in open quantum systems \cite{Alickiwork,Lindbladbook,alicki2004thermodynamics},  no work is extractable from the system: conversely, only the piston can do work on the system. This rule can be proved to be strictly obeyed \textit{if the bath-induced evolution is Markovian} (App.).  Yet,   our analytical results  and  numerical simulations (Fig.1) show  that while the system interacts with a  bath on non-Markovian time scales,  \emph{net work can be performed in a cycle by the system, i.e.,  the piston can be coherently amplified.}
Namely, under non-Markovian dynamics, we can  ensure $W^{ext}_{cycle}>0$ in Eq. \eqref{eq:energypart} by  choosing $\dot{\omega}(t)$ to oscillate \textit{out of phase}  with $s(t)$. For weak S-B coupling and $T=0$ we will explicitly show that $W^{ext}_{cycle}>0$ is indeed possible only in the non-Markovian limit on the modulation rate, $\Omega \gg \frac{1}{t_c}$. More elaborate analysis, whereby the piston field ``dresses'' the qubit states with which the bath interacts\cite{dressedstates,*kofPRA96}, yields qualitatively similar results.

We wish to evaluate the work performed by a qubit in contact with a bath at temperature T after a non-selective measurement of its energy, over a period (cycle) of its Stark-shift modulation of the form 

\begin{equation}
\omega(t)=\omega_a+ \delta Sin\Omega t. 
\label{eq:omega}
\end{equation}

\noindent To this end we shall use the results of the weak-coupling, non-Markovian master equation\cite{njp,kofmantemp,shahmoonPRA13}, whereby

\begin{equation}
s(t)=e^{-J(t)}(\int\Delta R(t')e^{J(t')}dt'+s(0)).
\label{eq:s(t)}
\end{equation}

\noindent Here the relaxation integrals  $J(t)=J_{g}(t)+J_{e}(t)$ , $J_{g(e)}(t)=\int_{0}^{t}R_{g(e)}(t')dt'$ and
$\Delta R(t)=\frac{1}{2}(R_{g}(t)-R_{e}(t))$, depend on the  effect of the  non-Markovian (B) bath: these are the  bath-induced transition rates $R_e(t)$ ($|e\rangle \mapsto |g\rangle$) and $R_g(t)$ ($|g\rangle \mapsto |e\rangle$). Both $J(t)$ and $R(t)$ are partly oscillatory on non-Markovian time scales, reflecting the partial reversibility of S-B dynamics. They are proportional to the square of the system-bath  coupling strength, $\eta$.

Since the coupling is assumed weak, $s(t)$ can be expanded
as
$
s(t)  \approx  s(0)(1-J(t))+\Delta J(t)+O(\eta^{3}),
$
where $\Delta J(t)=\int_0^t\Delta R(t')dt'$.
The first term in the expansion, $s(0)$, does not contribute to the
work ($\int_0^{\frac{2\pi}{\Omega}} s(0)cos\Omega tdt=0$). The other terms  are $O(\eta^2)$, so we set $s(0)\approx-1/2$, at $T \approx 0$.
The universal formula \cite{njp,kofmantemp} yields the relaxation integrals $J(t)$ and $\Delta J(t)$ through   $J_{g(e)}(t)$ as the spectral overlap of the bath response $G_T(\omega)$ and the modulation spectrum $F_t(\omega):$ 

\begin{gather}
J_{e(g)}(t)=\int_{-\infty}^{\infty}d\omega G_{T}(\omega)F_{t}(\omega_a \mp \omega)= \notag\\
\frac{1}{2\pi}\int_{-\infty}^{\infty}d\omega G_{T}(\omega)|\int_{0}^{t}dt'e^{i(\omega +\omega_a )t'}\epsilon(t')|^{2},
\label{eq:J}
\end{gather}

\noindent where the modulated phase factor $\epsilon(t)=e^{i\int_{0}^{t}dt'\omega(t')}$.
Using the Bessel expansion of $\epsilon(t)$ for our $\omega(t):$ $e^{-i\frac{\delta}{\Omega}Cos\Omega t}=\sum_{n=-\infty}^{\infty}i^{n}J_{n}(-\frac{\delta}{\Omega})e^{in\Omega t}$,
and  assuming a weak modulation, $\frac{\delta}{\Omega}\ll1$, the  expression for work in a cycle  reduces to 

\begin{eqnarray}
W^{ext}_{cycle}\approx\frac{\delta}{2\pi}\int_{-\infty}^{\infty}G_{0}(\omega)\frac{2\pi}{(\omega^+)^{2}} \times \nonumber\\
\left(sinc(\frac{2\pi}{\Omega}(\omega^+ +\Omega))+sinc(\frac{2\pi}{\Omega}(\omega^+-\Omega))\right)d\omega
\label{eq:wfin}
\end{eqnarray}

\noindent where  $G_0(\omega)$ is the zero-temperature bath response, and $\omega^+=\omega+\omega_a$. This expression may be approximated as

\begin{equation}
W^{ext}_{cycle}\approx- \delta \int_{0}^{\frac{2\pi}{\Omega}}J_{g}(t)\Omega cos\Omega t dt.
\label{eq:wapprox}
\end{equation}

\noindent  It shows that in the strongly non-Markovian limit $\Omega\gg \delta \sim \omega_a,$  the sign of $W^{ext}_{cycle}$ oscillates with $\Omega$, for a fixed $\omega_a$, and thus allows for \textit{either positive or negative work extraction}, as opposed to the Markovian limit (App.). Hence the work invested by the NSM (Eq. \eqref{eq:workin}) can be partly extracted in a non-Markovian cycle.

We next clarify how    S can   regain  the  energy deposited by the measurement, using our analytical results  and simulations (Fig. 1--main panel): \textit{The source of work is seen  to be only  the change of  $\langle H_{SB}\rangle $, the system-bath correlation energy.  }
       The rapid variation of the extractable work  with the cycle duration $t_{cycle}$ (Fig. 1a) proves that work retrieval from $\langle H_{SB}\rangle $  is \textit{limited to non-Markovian time scales }: $t_{cycle}=\frac{2\pi}{\Omega}$ should  be   shorter than  $t_c$,  the bath memory time,  to ensure  work performance by the system enabled by an unread QND measurement.   
   The reason for this anomalous effect is the  (partly reversible) S+B dynamics  expressed by the oscillatory relaxation integrals $J_g(t)$ and $J_e(t)$  on non-Markovian time scales triggered by the measurement.

  \begin{figure}[htbp]
	\centering
				\includegraphics[width=0.5\textwidth]{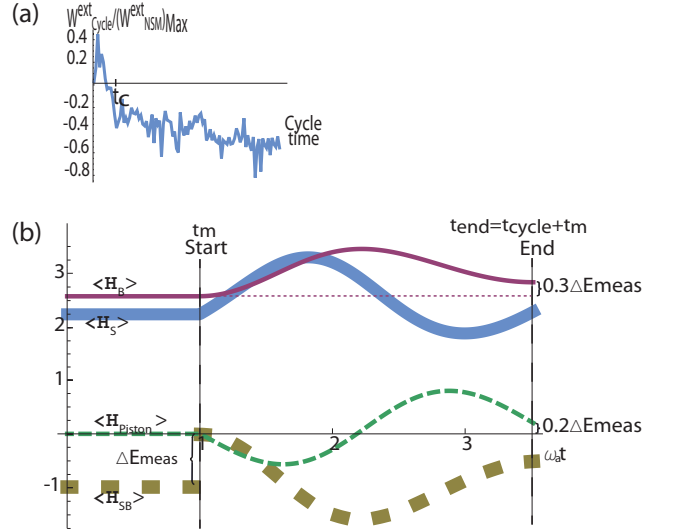}
		\caption{Work extraction by  measurements. Main panel: Simulations of the first-cycle evolution of the energy of the system (solid thick blue line), the  bath (solid thin purple line), the classical piston (dashed thin green line), and the system-bath correlations (dashed thick brown line). The piston periodically  modulates  the  TLS frequency, $\omega(t)=\omega_a+\delta sin\Omega t$.  
The  period of the modulation starts with an unread measurement of the TLS energy.  The parameters for the  curves are  $\omega_a=1$, $\delta=1/4$, $\Omega=5/2$. The S-B coupling spectrum is a Lorentzian centered at $\omega_0$ of width $t_c^{-1}$ (inverse correlation or memory time of the bath), with  $\omega_0-\omega_a=3/7$ , $t_c=10$ and the inverse bath temperature is $\beta=3.74$. The cycle starts with a non-selective measurement at time $\omega_a t_m=1$  and ends at $\omega_a t_{end}=3.51$. The measurement invests $\Delta E_{meas}=-\langle H_{SB}(t=t_m)\rangle $ in the system. Here the piston $\langle H_{Piston}\rangle $ is seen to have gained energy during the cycle, but not at the expense of $\langle H_S\rangle $ which returns to its initial value. Since $\langle H_B\rangle $ has also gained energy,  the source of work is  the change in $\langle H_{SB}\rangle $, the system-bath correlation. The simulations imply that $W^{ext}_{tot}=W^{ext}_{cycle}-\Delta E_{meas}<0$,  although $W^{ext}_{cycle}=\langle H_{Piston}(t_{end})\rangle-\langle H_{Piston}(t_m)\rangle >0$. Inset: extractable work in a cycle $W^{ext}_{cycle}$ normalized to  the maximum $(W^{ext}_{NSM})_{Max}$  as a function of the cycle duration $t_{cycle}$. It is seen that $W^{ext}_{cycle}>0$ (work done by the system) requires $t_{cycle} \lesssim t_c$, $t_c$ being the bath memory time. Same parameters as in main panel 1.
}
\label{fig:energies}
\end{figure}

 \textbf{ IV The revised work-information relation }

 We are now in a position to address the fundamental questions that motivate this paper:  What is the difference between the maximal work extracted in a cycle via a selective (read) measurement, $(W^{ext}_{sel})_{Max}$, and  its non-selective (unread) counterpart, $(W^{ext}_{NSM})_{Max}$? How do they differ from their SL counterparts?

Let us  define the measurement basis  as $|j\rangle, \hspace{2mm}{j=e,g}$. Then $p_j$ is the probability of finding the state $\rho_{S+B}$ in the state j, and $\rho_{S+B}^j$ is the state of the supersystem after the measurement. The maximum extractable work by a nonselective measurement  is given by Eq. \eqref{eq:main}. Its counterpart for selective measurement is $(W^{ext}_{sel})_{Max}=\sum p_{j}\Delta E_{j}-T\sum p_{j}\Delta \mathscr{S}_{j}$.  where $\Delta E_j$ and $\Delta \mathscr{S}_j$ are the respective changes when the state j is measured, and $\Delta E_{meas}=\sum_j p_j \Delta E_j$.
The difference between work extraction based on selective and nonselective measurement is

\begin{gather}
(W_{sel}^{ext})_{Max}-(W_{NSM}^{ext})_{Max}= \nonumber \\
T(\mathscr{S}(\sum_jp_j\rho^j_{S+B})-\sum_j p_j\mathscr{S}(\rho^j_{S+B}))=
T \mathscr{H}(\{p_j\}) \nonumber \\
=T \mathscr{S}(\rho_S)
\label{eq:workdif}
\end{gather}
The last step follows from the equality of the Shannon entropy $\mathscr{H}$ of the  system  and its von Neumann entropy $\mathscr{S}(\rho_S)$ when  $\rho_S$ is diagonal, as in our case, if we set $k_B=1$ and $\mathscr{H}=ln2H_{Shannon}$,  $H_{Shannon}$ being the standard definition of the Shannon entropy. Explicitly, in our case $\mathscr{H}=-plnp-(1-p)ln(1-p)$. The foregoing expression may be recast in the form

\begin{equation}
(W^{ext}_{sel})_{Max}=(W^{ext}_{NSM})_{Max}+ 
W_{SL}.
\label{eq:selvsnonsel}
\end{equation}

\noindent Here the first term on the r.h.s. is Eq. \eqref{eq:main} and  $W_{SL}=T\mathscr{H}(p)$ denotes the standard SL work extraction:  the energy required  to reset the probe to its
initial state.

 Equation \eqref{eq:selvsnonsel}  is our main result: it implies that the maximal  \textit{work extractable via a selective  measurement is higher than what is expected from the  SL relation} between work and information:  the S-B correlations  increase this extractable work  by  $(W^{ext}_{NSM})_{Max}$.  In the standard (Markovian) case, where the system-bath interaction energy is assumed negligible, $(W^{ext}_{NSM})_{Max}=0$ and $(W^{ext}_{sel})_{Max}=T\mathscr{H}(p)$. Hence, in the Markovian case,  the selective   measurement (in this scenario) does not yield any work  beyond the Landauer cost\cite{landauer} $W_{SL}=T\mathscr{H}(p)$ of resetting (``cleaning'') the memory of the probe after each measurement. In the non-Markovian case, the measurement  process requires higher investment of work, since it changes the  system-bath correlation energy, but also allows more  work to be extracted than in the Markovian case.   
 
\begin{figure}[htbp]
\centering
					\includegraphics{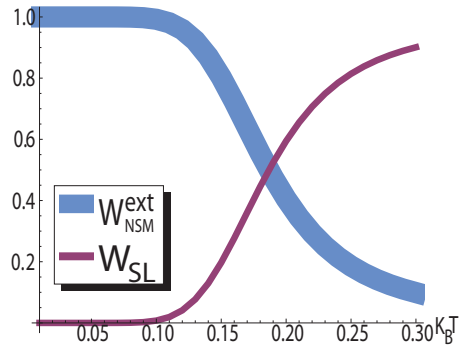}
				\caption{The Szilard-Landauer work $W_{SL}$ (thin purple) and  $(W_{NSM}^{ext})_{Max}$ (solid blue), both normalized to $(W_{sel}^{ext})_{Max}$ (Eq. \eqref{eq:selvsnonsel}),  as a function of the single-bath temperature. Even at T=0, where Szilard-Landauer work vanishes,  work can be extracted via a measurement in the non-Markovian time-domain. 
}
\label{fig:wselvswnsl}
\end{figure}

The NSM effect discussed above  allows work extraction at $T=0$ and \textit{without any information gain}: both features take us beyond  the SL \cite{Szilard,landauer} information-work balance or its IT extensions\cite{uedatheory,JacobsPRE2012}. Namely, even at $T=0$,  $\langle H_{SB}\rangle_{Eq}<0$ and its change by $\Delta E_{meas}$ powers the cycle, yielding  $(W_{sel}^{ext})_{Max}=(W_{NSM}^{ext})_{Max}>0$ (Fig. \ref{fig:wselvswnsl}). The entire work then originates from the non-Markovian change of system-bath correlations. 

Whereas $W_{NSM}^{ext}$ has been shown above to exceed the SL bound, it can be argued that $W_{sel}^{ext}$ can be even higher. To this end, consider that in the case of selective measurements, the work extraction will be the weighted sum of that obtained by each measurement result 

\begin{eqnarray}
W_{sel}^{ext}= 
\rho_{ee}(0)W_e+\rho_{gg}(0)W_g= \nonumber \\
\oint [J_g(t) \rho_{gg}(0) \dot{\omega}_g(t)-J_e(t) \rho_{ee}(0) \dot{\omega}_e(t)]dt
\label{eq:wtwo}
\end{eqnarray}

\noindent where the two modulations $\dot{\omega}_{e(g)}(t)$ may be different. Suppose we choose $\dot{\omega}_{e(g)}$ so that  they maximize the total work $W_{sel}^{ext}=(W_{sel}^{ext})_{Max}$. If the corresponding modulations happen to coincide, $(\dot{\omega}_{e}(t)_{Max}=\dot{\omega}_{g}(t)_{Max}$), then the resulting expression is the same as $(W_{NSM}^{ext})_{Max}$ (Eq. \eqref{eq:energypart}). If, on the 
contrary, $(\dot{\omega}_{e}(t)_{Max}\neq \dot{\omega}_{g}(t))_{Max}$, then the two bounds differ. Clearly, we may then have 

\begin{equation}
(W_{sel}^{ext})_{Max}>(W_{NSM}^{ext})_{Max},
\label{eq:selvsnosel}
\end{equation}
 since by choosing $\dot{\omega}_{e}(t)$ and 
$\dot{\omega}_{g}(t)$ to be \textit{out of phase } throughout the cycle, the two terms in Eq. \eqref{eq:wtwo}  acquire the same sign, i.e.,  add up.  By contrast, in 

\begin{equation}
W_{NSM}^{ext}=\oint [J_g(t) \rho_{gg}(0) -J_e(t) \rho_{ee}(0))] \dot{\omega}(t)dt,
\label{eq:wns}
\end{equation}
 
 the two terms have opposite signs, so that $(W_{sel}^{ext})_{Max}$ can exceed $(W_{NSM}^{ext})_{Max}$, q.e.d.

\textbf{V Consistency with the second law}

The second law is  upheld  (in the sense that perpetual motion\cite{niewuallah} becomes forbidden) only when  we account for the \textit{energy and entropy cost} of  changing the ``supersystem'' state $\rho_{S+B}$ from its correlated  Gibbs form at equilibrium $\rho_{Eq}$  to  its post-measurement  product-state form $\rho_S \otimes \rho_B$. This cost becomes evident only in a description of the evolution in terms of the \textit{total} Hamiltonian  and the corresponding state $\rho_{tot}$ that encompass the degrees of freedom of the \textit{probe and the supersystem}, P+S+B. Explicitly,

\begin{equation}
H_{tot}=H_0+H_{BP}+H_{SP},
\label{eq:htot}
\end{equation}

 \noindent where  $H_{SP}$ is the system-probe coupling term, $H_{BP}$ allows a fast decorrelation between the system and probe, and $H_0$ describes the ``supersystem'', system+bath.
Using the fact that the total Hamiltonian is cyclic, $H_{tot}(\tau)=H_{tot}(0)$, the total extractable work in a cycle (by P+S+B) can be written as

\begin{gather}
W^{ext}_{tot}(\tau)=-\int_0^{\tau}
Tr \left[U(t)\rho_{tot}(0)U^{\dagger}(t) \right]\dot{H}_{tot}(t)dt= \notag \\
-Tr \left[\rho_{tot}(\tau)H_{tot}(\tau)-\rho_{tot}(0)H_{tot}(0)\right] \notag \\
+
\int_0^{\tau}
Tr\left[ \dot{\rho}_{tot}(t)H_{tot}(t)\right]dt
\label{eq:nonnepro}
\end{gather}

Upon inserting  the expression for $ \dot{\rho}_{tot}(t)$ and calculating the trace we find that the second term on the RHS is zero. The first term thus represents the energy change of the supersystem. Since initially the supersystem and the probe were at thermal equilibrium, any energy change should be \textit{positive}.
We then find:

\begin{gather}
W^{ext}_{tot}(\tau)= \notag \\
-Tr \left[U(\tau)\rho_{tot}(0)U^{\dagger}(\tau)H_{tot}(0)-\rho_{tot}(0)H_{tot}(0)\right].
\end{gather}
 Here the first term is the final mean energy and the second is the initial one.
 
  Because the total dynamics is unitary, the entropy of $\rho_{tot}$ is fixed. This implies that  the final mean energy (first term) must be greater or equal to the initial one (second term), as the thermal-equilibrium initial state minimizes the mean energy at fixed entropy.

\begin{gather}
W^{ext}_{tot}= \notag \\
-\oint Tr\{\rho_{tot}\dot{H}_{tot}\}dt=
-\Delta E_{meas}+W^{ext}_{cycle}<0.
\label{eq:genwork}
\end{gather}

\noindent The negativity of Eq. \eqref{eq:genwork}  under a cyclic \textit{unitary} evolution of the total Hamiltonian, starting from equilibrium   of the supersystem and the probe, can be proved completely  generally. It shows that the second law that forbids drawing work from a single bath only applies to \textit{the entangled }evolution of S+B+P and that  their standard separability assumption\cite{Szilard,landauer,caves,scullyPRL2001,uedanature,segal2010,MaruyamaRMP2009,Ruschhauptjphisb06}  fails for sufficiently fast cycles.

\textbf{VI Multiple cycles: Resetting cost for non-selective measurement?} 

It might be suspected  that   resetting (purifying)  the probe is necessary  if we wish to reuse it  in successive cycles and that would add to the thermodynamic cost  as per the SL principle \cite{landauer}. 
Clearly, the  probe  cannot  circumvent the SL resetting cost $W_{SL}$  when
  \textit{selective} measurements are required (Eq. \eqref{eq:selvsnonsel}). 
  Yet, this \textit{is not the case} for a NSM, which requires \textit{no resetting} because the   bath  can rapidly decorrelate \textit{P} and the supersystem  following the measurement (Eq. \eqref{eq:proc}), but prior to the next cycle: 

\begin{equation}
\rho_{tot}\mapsto \rho_{S+P} \otimes \rho_{B}\mapsto \rho'_P \otimes \rho'_{S+B}.
\label{eq:rhod}
\end{equation}

 The absence of resetting cost after a NSM   follows from a remarkable observation:   \textit{a single probe qubit has the same NSM effect on any number of system cycles}. This holds since  it does not matter how each cycle   changes the 
state of the probe, $\rho_{P}$, because the resulting state commutes
with the probe's  $\sigma^P_{z}$ (Eq. (4)). In particular, work is extractable even if \textit{ P} is in   \textit{
the fully mixed (infinite-temperature) state:} for $\rho_{P}=\frac{1}{2}I^P$,  the
CNOT leaves the probe qubit unchanged, i.e., it cannot be read out and yet the \textit{same} probe qubit can still perform the required NSM on   the system qubit as often as we
like, i.e.  our probe is never used up.  This is because  Eq. \eqref{eq:rhod} still holds in this case: S+P become correlated by the measurement (after the previously described fast modulation period), but then  the correlations between the system and the probe decay   via thermal relaxation and revert to  a product state (Fig. 3). After this relaxation   the probe can be reused in the next cycle and  have  the same effect  as in the first cycle. Hence, \textit{there is no need of resetting the probe for further use in consecutive cycles, provided it performs repeated NSM.}

\begin{figure}[htbp]
	\centering
		\scalebox{0.4}{\includegraphics[width=1\textwidth]{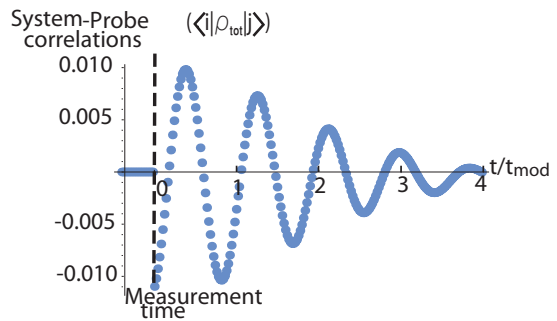}}
	\caption{The role of probe-system correlations and their destruction by the bath.  Decay of the off-diagonal system-probe (S-P) elements (correlations) $\left<i| \rho_{tot}|j\right>$ under $H_{tot}$ (Eq. 1) where $\left|i\right>=\left|e,1,n\right>$ and $\left|j\right>=\left|g,0,n\pm1\right>$, the entries denoting the system, probe and bath quantum numbers, respectively. The parameters are the same as in Fig. 1. The probe frequency is 10/7 $\omega_a$. The decay time of the correlations is that of the oscillations envelope, here $\sim 4$ modulation periods $(4t_{mod})$. After this time the probe can be reused for the next cycle. 
	 } 
	\label{detsyscor}
\end{figure}

\begin{figure}
	\centering
		\includegraphics{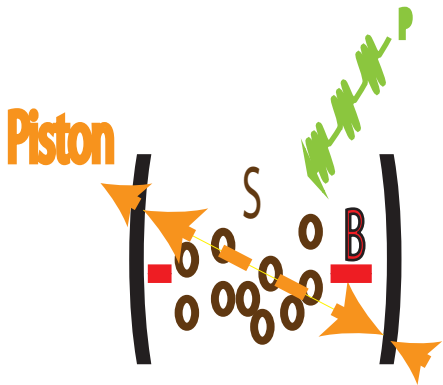}
	\caption{Possible experimental setup: The measured atomic TLS (S) in a  cavity bath (B) does work on a frequency-modulating piston mode  of the cavity . The time-modulated  probe   effects brief, non selective, QND measurements.}
	\label{fig:exp}
\end{figure}

\textbf{VII Experimental Scenario}

A feasible experimental test of these predictions may involve (Fig. \ref{fig:exp}) the following ingredients:

 A) An an ensemble of N two-level atoms (S) in a   cavity (resonator) of length L, whose \textit{high-Q field modes} that are near-resonant with the atomic resonance frequency $\omega_a$ constitute the non-Markovian intracavity bath (B) with memory time  $t_c \sim \frac{LQ}{c}$: $t_c \sim 10^{-4} sec$ are feasible at present \cite{PetrosyanPRA09,nvcenter}.  The interaction energy $\langle H_{SB} \rangle$ and thus the NSM induced work (Eqs. \eqref{eq:workin}, \eqref{eq:main}) may attain the GHz range, as they scale with the collectively-enhanced N-atom coupling to the bath, $\sqrt{N}$.

B) An off-resonant, coherent (classical) signal constitutes the piston  with amplitude  $E_0(1+p Cos\Omega t)$, ($p<1$) that modulates the atom level-distance $\omega _(t)$ at a rate $\Omega\gg 1/t_c$ by periodic Stark shift.  

C) Injected pulses, much shorter than $t_c$, can   probe the atomic-state population in a QND fashion,  on a sub-nsec/psec timescale.  Specifically, the CNOT protocol of QND measurements in Eqs \eqref{eq:totham}-\eqref{eq:measrhotot} demands entangling the polarization of a probe photon with the magnetic sublevels of one of the TLS states. If this photon is to be reused, the photon-atom states are to be disentangled by a depolarizing environment as per Sec. V. Yet even a classical probe may effect a QND measurement by its polarization action  on the atomic sublevel population\cite{ScullyBOOK97}. As discussed in Sec VI, NSMs may be performed by a probe that is  \textit{arbitrarily} noisy in its polarization. 

Both the weak-modulation (Eqs. \eqref{eq:omega} and  \eqref{eq:s(t)} ) and the more elaborate \textit{exact solution} of this \textit{S+B} model \cite{kofPRA96}  predict that $W_{NSM}^{ext}$ (Eqs.\eqref{eq:wapprox}) and $W_{sel}^{ext}$ (Eqs. \eqref{eq:wtwo})  can  extract work via NSM at $T=0$ (an \textit{empty cavity}, or field-bath vacuum and for \textit{ground-state} atoms) only  if the cycle duration is much shorter than $t_c$. The extracted work will be manifest by the \textit{amplification (lasing)} of the off-resonant coherent piston  mode despite the absence of atomic population-inversion or bath heat energy, at the expense of the  S-B interaction (correlation) energy (Eq.\eqref{eq:workin}). The distinctive signature of this amplification is that it is restricted to $t_{cycle}=\frac{2\pi}{\Omega}\leq t_c:$ as $t_{cycle}$ starts exceeding $t_c$, amplification will  revert to loss. 
 The described process is akin to intracavity \textit{parametric conversion of external driving} (probe pulses), resulting in signal (``piston'' mode)  amplification \cite{Scully2011,ScovilPRL59}, but  it is unique in its reliance on  system-bath correlations, and in its insensitivity to the probe noise.

\textbf{VII Discussion}

We have shown the  possibility of extracting useful work from  an \textit{open} quantum system following either a non-selective (unread) QND measurement (NSM) (Eq. \eqref{eq:wapprox}) or a selective (read) measurement (SM) (Eq. \eqref{eq:wtwo}). In both cases, a modulator (piston) can take work and gain energy from the system (be coherently-amplified) in a closed cycle. This work originates  neither from the probe free-energy\cite{uedatheory,JacobsPRE2012}  nor from the heat energy of the  bath  (as in Szilard's engine\cite{Szilard,landauer,caves,scullyPRL2001,uedanature,segal2010,MaruyamaRMP2009,Ruschhauptjphisb06}) but from a hitherto unexploited (and little-discussed) source: the \textit{inevitable} change of the  system-bath (S-B) correlation (interaction) energy (see \cite{niewuallah}) by a \textit{brief}  QND measurement  \cite{nature,chapinnat08,njp,njp10,Gonzalo,Jahnke}.
Only  \textit{non-Markovian} supersystem  (S+B) dynamics can yield  extractable work  following such a measurement, as opposed to its   Markovian limit that ignores system-bath correlations (Fig. 1 a).

When discussing these effects, certain misunderstandings must be dispelled: (i) The proposed   work resource \textit{cannot} be explained   by viewing either the unread probe or the piston as  a fictitious additional   ``bath''. Neither constitutes a proper heat bath: the piston is a zero-temperature and zero-entropy classical drive that only gains work and energy from the system, while  \textit{the probe must act impulsively}, unlike usual sources of noise of heat. (ii) Nor can one deny the cycle is triggered by a measurement: even if the measurement is unread, \textit{it is still a measurement}, as evidenced by the S-P correlations (see Eq. \eqref{eq:rhod}).  (iii) Recently considered measurement-cost  (Markovian) effects \cite{uedatheory,JacobsPRE2012} are beyond the scope of our scenario (see SI2).

The colloquial maxim ``there are no free lunches'' applies to the predicted effect, i.e., the surplus work is  allowed only by extra investment of energy  consistently with the first law (otherwise it would enable a ``perpetuum mobile'' machine)\cite{niewuallah,FordPRL06} and the second law is also upheld (Eq.\eqref{eq:genwork}). Yet this effect may allow us to study the possibilities of transforming  energy input (e.g., electromagnetic probe pulses which may be \textit{very noisy} as argued in Sec. VI, similarly to \cite{Scully2011}.) into \textit{useful work} [coherent signal (piston) amplification] via rapid modulations of thermalized quantum systems

The present engine model, in which the system is always coupled to a single  bath and yet may perform useful work, is potentially important for  systems  totally embedded in  a single bath, such as a cavity,  so that  conventional heat-engine  (two-bath) thermodynamic cycles may be impossible to implement.
  Further investigation may include  brief disturbances other than measurements, e.g., phase flips  of a TLS in a  bath \cite{durgaphysreva}.

This research was supported by DIP, ISF, BSF and CONACYT.

\setcounter{equation}{0}
\renewcommand{\theequation}{M\arabic{equation}}

\section*{Appendix}
\setcounter{equation}{0}
\renewcommand{\theequation}{A\arabic{equation}}

\textbf{1 Measurement cost  in our scenario compared to Maxwell's demon models}

A) \textit{Measurement cost for asymmetric detector}

The SL principle has been extended  to a  detector/memory modeled by a quantum particle in an asymmetric double-well potential (the Sagawa-Ueda (SU) model\cite{uedatheory}). Such asymmetry may reduce the expense of resetting the detector to its initial state at the cost of increasing the work required to perform the measurement, but the sum of the two costs remains the \textit{same as the  cost set by the SL principle}. In that case, the mean energy of the
device (D) may be altered by the measurement, and in particular
the device may exchange energy with the measured system. For such a device the resetting cost may  differ from 
Landauer's. SU assume a total Hamiltonian of the form:

\begin{gather}
H(t)=H_{D}(t)+H_{B}+H_{DB}(t) ;\;  \nonumber \\
H_{D}(0)=H_{D}(\tau)=H_{D};\; H_{DB}(0)=H_{DB}(\tau)=0
\label{eq:H(t)}
\end{gather}

The SU Hamiltonian does not describe the measurement itself, only the
detector-bath interaction $H_{DB}(t)$. By contrast,  in our scenario the $H_{SD}(t)$ term in Eq. (1) causes a change of $\langle H_{SB} \rangle$ a S+B correlation term, which is missing from the SU analysis: they adopt the S+B separability paradigm, whereas we do not.
Thus, in the SU model, the measurement conserves $\langle H_{SB} \rangle +\langle H_{D} \rangle$, as opposed to our dynamics, hence the difference in measurement cost and post-measurement work extraction.
In our case the final state of the D has the same energy as  its initial
state. Yet our measurement cost is given by Eq. (2) and is nonzero due to S+B correlation change.

Similarly, the work invested in the measurement in our scenario  is performed not on D, as in the SU model, but on S+B, and is equal to the change in $\langle H_{SB} \rangle$ in Eq. \eqref{eq:enepm}.

In the SU  model, the initial state of D is assumed to
be a thermal state corresponding to a particular subspace, $\mathbf{H}_{0}^{D}$
of its total  Hilbert space , whereas the final  state of D
may  have changed. This leads to the following expression for the measurement cost in the SU model, which is  the
change in the free energy of D due to the measurement process

\begin{equation}
\Delta F_D\equiv \Sigma_{k}p_{k}F_{k D}-F_{0D}
\label{eq:F}
\end{equation}

where $p_{k}$ is the post-measurement probability for D to be found in
subspace $\mathbf{H}_{k}^{D}$. By contrast,  in our case the final state of the D has the same energy as  its initial
state  and  $p_{0}=1$, giving: $\Delta F_{D}=0$. Yet our measurement cost is given by Eq. \eqref{eq:workin} and is nonzero due to S+B correlation change.

Similarly, the work performed on D by the measurement  is assumed by SU \cite{uedatheory}
to be the negative of the change in energy of D, which in our scenario 

\begin{equation}
W_{meas}^{D}\equiv Tr\left\{ (H_{D}+H_{B})\left.\rho_{DB}\right|_{0}^{\tau}\right\} =0
\label{eq:SUWD}
\end{equation}

Namely,  in our scenario this change in the  mean energy of D  vanishes. By contrast, we have post-measured work extraction due to the change in S+B correlations by a measurement after time $\tau$:

\begin{equation}
W^{ext}=\langle H_{S+B}(\tau)\rangle_D-\langle H_{S+B}(0)\rangle_D\neq W^D_{meas}=0.
\label{eq:Wext}
\end{equation}

b) \textit{Measurement cost for degenerate detector}

In Ref \cite{JacobsPRE2012} a detector \textit{D} consisting of \textit{degenerate} states, initially in thermal equilibrium, performs a selective measurement (SM) on \textit{S} in order to extract work in a cycle as Maxwell's demon. Since a SM reduces the entropy of \textit{S}, the entropy of \textit{D} must correspondingly rise. Hence, D must not be in a maximal entropy state prior to the measurement. Neither can its temperature be the same as that of \textit{S}, i.e., \textit{T}.  The required lowering of the temperature and entropy of \textit{D} are achieved by isothermally (quasistatically) lifting the degeneracy of the levels of \textit{D} at a cost $\Delta E_D$. The maximal work extraction by \textit{D} is \cite{JacobsPRE2012} the free energy lost by \textit{D} 

\begin{equation}
W_D= \Delta E_D +T \Delta \mathscr{S}
\label{eq:wjac}
\end{equation}

Although \eqref{eq:wjac} superficially looks \textit{similar} to our Eq. (9), it is essentially different in that $\Delta E_D$ is determined by the detector \textit{temperature}, which is \textit{irrelevant} for the impulsive NSM used in our scenario to change $\langle H_{SB} \rangle$. In particular, $\Delta E_D=W_D=0$ at $T=0$ in Ref. \cite{JacobsPRE2012}, as opposed to our Eqs. (3) and (9), where the measurement cost does not change  $\langle H_D \rangle=0$, but $\langle H_{SB}\rangle$ changes even at $T=0$. The temperature and entropy restrictions on \textit{D} in Ref. \cite{JacobsPRE2012} do not exist in our model (see Sec. VIII). The reason for these differences is that by venturing beyond the \textit{S-B} separability paradigm we enable the entropy of \textit{S} to be reduced at the expense of $\Delta \mathscr{S}_{S+B}$ reflecting \textit{S-B} correlation change by the measurement, whereas in Ref. \cite{JacobsPRE2012} the bath \textit{B} does not affect the entropy balance during the measurement under the \textit{S-B} separability paradigm.

\textbf{2 Maximal work in a post-measurement cycle}

The post-measured S+B  supersystem is thus in a nonequilibrium state  that
can be harnessed to perform work on its way back to equilibrium.
The maximal work possible  is extractable in a cycle that is thermodynamically  reversible
apart from the measurement ``stroke''\cite{maxwork}. Were  $\rho'_{S+B}$ a thermal (Gibbs) state (for \emph{some} temperature),
we could use standard processes\cite{alicki2004thermodynamics} to {}``close the cycle'' by a reversible process, and 
the maximal extractable work would then  be given by the difference in the Helmholtz
free energy between $\rho'_{S+B}$  and the original equilibrium state\cite{ScullyBOOK97}. However, since $\rho'_{S+B}$ 
is not a Gibbs state, it is not clear that this upper bound on work is appropriate.

To find a thermodynamically reversible process  that would bring the post-measured  state back to equilibrium, we resort to a \textit{nonstandard} procedure that allows maximal work extraction. Namely,  we envision that the supersystem $S+B$ is embedded in a Markovian bath $B_M$, \textit{at the same temperature} as B, $T=\frac{1}{\beta}$. 
The supersystem  $S+B$ equilibrates with $B_M$ at time $t_{Eq}$, say via coupling between B and $B_M$.  Since $B_M$ is Markovian we can neglect its correlation with $S+B$. Yet the correlations between S and B persist 
much longer, because B is non-Markovian, with correlation (memory) time $t_c \gg t_{Eq}$.

\begin{figure}
	\centering
		\includegraphics[width=0.5\textwidth]{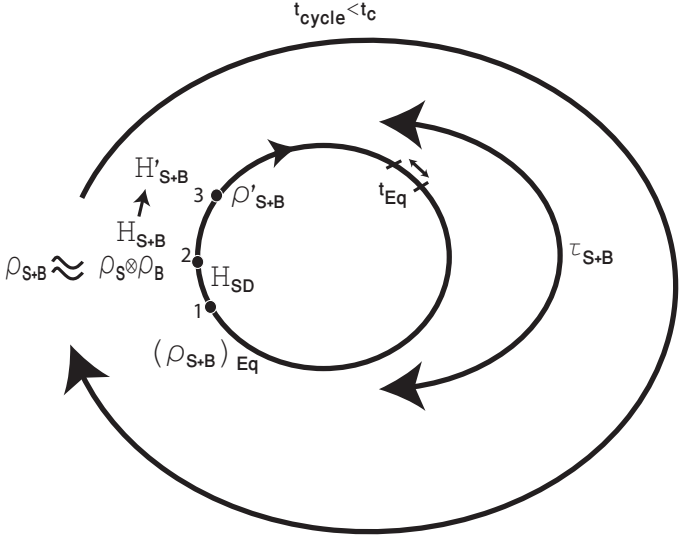}
		\caption{Work extraction by measurements from S-B correlations:  Optimal cycle that consists of 3 stages (see text): 1- measurement, 2-stabilization, 3-modulation.}
	\label{fig:swplane}
	\end{figure}

The stages of this nonstandard, optimal  cycle are as follows (Fig. \ref{fig:swplane}):
(1) The initial equilibrium state $\rho_{B_M}\otimes \rho_{SB}$, where $\rho_{SB}= \rho_{Eq}=\frac{e^{-\beta H_{SB}}}{Z}$, undergoes at time $t=0$ a measurement of \textit{S} (Eqs. \eqref{eq:enepm} and \eqref{eq:workin})  that leaves $S+B+B_M$ in (approximately) the product state   $\rho_{B_M}\otimes \rho_{S}\otimes \rho_{B}$.
(2) We next stabilize $\rho_S \otimes \rho_B$ by making a sudden change of  the \textit{S+B} Hamiltonian: $H_{S+B} \rightarrow H'_{S+B}$, so  that the overall state becomes $\rho_{B_M}\otimes \frac{e^{-\beta H'_{S+B}}}{Z'}$.  The change of work is $W_{stab}= \langle H_{S+B}' \rangle-\langle H_{S+B}\rangle$. We are guaranteed that such stabilization is possible \cite{maxwork,Takara201088}, but it may not be feasible if we only act on \textit{S} (by modulating the qubit level-distance).
(3) Subsequently, we change $H'_{S+B} \rightarrow H_{S+B}$ by modulation over time $\tau_{S+B}\gg t_{Eq}$, i.e. quasistatically and \textit{isothermally} as concerns $B_M$,  until we attain the original equilibrium state $\rho_{B_M}\otimes \frac{e^{-\beta H_{S+B}}}{Z}$ and thereby close the cycle. The work change during the isothermal stage is $W_{isot}=\Delta E_{isot}-T\Delta S_{isot}$.

The overall optimal  cycle is described as follows: (i) In the first stroke, the energy \emph{cost} of the measurement is  (see Eq \eqref{eq:workin})
$
\Delta E_{meas}=\langle H\rangle_{\rho'}-\langle H\rangle_{\rho}
$.
The NSM  increases the VN
entropy: 
$
\Delta \mathscr{S}_{meas}=\mathscr{S}\left(\rho'\right)-\mathscr{S}\left(\rho\right).
$
 (ii)In the next (return) stroke, the stabilization (sudden)  Hamiltonian change implies that
work is performed \emph{by the system}:
$
W_{sudden}=\langle H\rangle_{\rho'}-\langle H'\rangle_{\rho'}
$
 and the entropy is unchanged. (iii) In the last stroke, the energy change of the supersystem is
$
\Delta E_{isotherm}=\langle H\rangle_{\rho}-\langle H'\rangle_{\rho'},
\hspace{0.3cm}
\Delta \mathscr{S}_{isotherm}=-\Delta \mathscr{S}_{meas}
$
and the extracted work during this stroke is \cite{alicki2004thermodynamics}:
$W_{isotherm}=-\Delta E_{isotherm}+T\Delta \mathscr{S}_{isotherm}.
$
(iv) Finally, combining these results for all strokes  one gets the expression in Eq.\eqref{eq:main} for $W_{extracted}=W_{sudden}+W_{isotherm}$

\textbf{3 No work can be extracted from a single Markovian-bath engine in a closed cycle}

We  consider  the  evolution of the TLS  state, $\rho_S(t)$, that is diagonal in the energy basis, with parametrically  time-dependent energy levels $E_e(t)-E_g(t)=\omega(t)$:

\begin{gather}
 {\dot \rho }_{ee}(t) = R_g(t)\rho_{gg}-R_e(t)\rho_{ee} \nonumber \\
 {\dot \rho }_{ee}(t) =-{\dot \rho }_{gg}(t)  
 \label{eq:tls}
\end{gather}

Let us now assume Markovian properties:
A)  $R_{g (e)}(t)\geq 0$;
B) Gibbs probability distribution in a stationary state (detailed thermal balance) at temperature $k_BT=\frac{1}{\beta}$

\begin{gather} 
R_{e}(t)\rho^{eq}_{ee}(t) = R_{g}(t)\rho^{eq}_{gg}(t)\ , \nonumber \\
\ \rho^{eq}_{jj}(t)= Z^{-1}(t)\exp\{-\beta E_j(t)\} \hspace{1cm} j\in (g,e)
\label{eq:detailedbalance}
\end{gather}

$Z(t)$ being the normalization constant.

To prove this result (which is consistent with known results) consider the following auxiliary expression
\begin{gather}
\sum_j {\dot \rho}_{jj} (\ln \rho_{jj} - \ln \rho^{eq}_{jj}) = \nonumber\\
 (R_{g}\rho_{gg} - R_{e}\rho_{ee}) \ln {\frac{\rho_{ee}}{\rho^{eq}_{ee}}}-
(R_{g}\rho_{gg} - R_{e}\rho_{ee}) \ln {\frac{\rho_{gg}}{\rho^{eq}_{gg}}}=\nonumber\\
R_{g}\rho^{eq}_{gg}(x\ln y-x\ln x+x-y)+ \nonumber\\
R_{e}\rho^{eq}_{ee}(y\ln x-y\ln y+y-x)\leq 0
\label{eq:unequality}
\end{gather}
where $x=\frac{\rho_{gg}}{\rho^{eq}_{gg}}$ and $y=\frac{\rho_{ee}}{\rho^{eq}_{ee}}$. Notice, that $R_{g}\rho^{eq}_{gg}(x-y)+R_{e}\rho^{eq}_{ee}(y-x)=0$ due to  assumption B).
The inequality in \eqref{eq:unequality} is obtained from the relation $a \ln a - a\ln b + b -a \geq 0$ (for $a,b \geq 0$) and  assumption A).
It implies the following inequality for the entropy  $S(t) = -k_B \sum_{j} \rho_{jj}(t)\ln \rho_{jj}(t)$:

\begin{gather}
{\dot S} = -k_B\sum_{j} {\dot \rho}_{jj} \ln \rho_{jj} \geq -k_B\sum_{j} {\dot \rho}_{jj} \ln \rho^{eq}_{jj} = {\frac{1}{T}} {\dot Q}\ ,\ \nonumber \\
 {\dot Q}= \sum_j {\dot \rho}_{jj} E_j
\label{eq:sdot}
\end{gather}

where we used the fact that $\sum_{j} {\dot \rho}_{jj} \ln Z(t)= \ln Z(t){\frac{d}{dt}}\sum_{j} \rho_{jj} = 0$.
\bigskip

{\it Since for a closed cycle the entropies and internal energies in the initial and final states of the system are equal, $W =   Q \leq  0$ (which is the second law of thermodynamics). This means that we cannot extract work from a single Markovian bath engine.}


\begin{thebibliography}{39}%
\makeatletter
\providecommand \@ifxundefined [1]{%
 \@ifx{#1\undefined}
}%
\providecommand \@ifnum [1]{%
 \ifnum #1\expandafter \@firstoftwo
 \else \expandafter \@secondoftwo
 \fi
}%
\providecommand \@ifx [1]{%
 \ifx #1\expandafter \@firstoftwo
 \else \expandafter \@secondoftwo
 \fi
}%
\providecommand \natexlab [1]{#1}%
\providecommand \enquote  [1]{``#1''}%
\providecommand \bibnamefont  [1]{#1}%
\providecommand \bibfnamefont [1]{#1}%
\providecommand \citenamefont [1]{#1}%
\providecommand \href@noop [0]{\@secondoftwo}%
\providecommand \href [0]{\begingroup \@sanitize@url \@href}%
\providecommand \@href[1]{\@@startlink{#1}\@@href}%
\providecommand \@@href[1]{\endgroup#1\@@endlink}%
\providecommand \@sanitize@url [0]{\catcode `\\12\catcode `\$12\catcode
  `\&12\catcode `\#12\catcode `\^12\catcode `\_12\catcode `\%12\relax}%
\providecommand \@@startlink[1]{}%
\providecommand \@@endlink[0]{}%
\providecommand \url  [0]{\begingroup\@sanitize@url \@url }%
\providecommand \@url [1]{\endgroup\@href {#1}{\urlprefix }}%
\providecommand \urlprefix  [0]{URL }%
\providecommand \Eprint [0]{\href }%
\providecommand \doibase [0]{http://dx.doi.org/}%
\providecommand \selectlanguage [0]{\@gobble}%
\providecommand \bibinfo  [0]{\@secondoftwo}%
\providecommand \bibfield  [0]{\@secondoftwo}%
\providecommand \translation [1]{[#1]}%
\providecommand \BibitemOpen [0]{}%
\providecommand \bibitemStop [0]{}%
\providecommand \bibitemNoStop [0]{.\EOS\space}%
\providecommand \EOS [0]{\spacefactor3000\relax}%
\providecommand \BibitemShut  [1]{\csname bibitem#1\endcsname}%
\let\auto@bib@innerbib\@empty
\bibitem [{\citenamefont {Szilard}(1929)}]{Szilard}%
  \BibitemOpen
  \bibfield  {author} {\bibinfo {author} {\bibfnamefont {L.}~\bibnamefont
  {Szilard}},\ }\href@noop {} {\bibfield  {journal} {\bibinfo  {journal}
  {Zeitschrift Physik}\ }\textbf {\bibinfo {volume} {53}},\ \bibinfo {pages}
  {840} (\bibinfo {year} {1929})}\BibitemShut {NoStop}%
\bibitem [{\citenamefont {Landauer}(1961)}]{landauer}%
  \BibitemOpen
  \bibfield  {author} {\bibinfo {author} {\bibfnamefont {R.}~\bibnamefont
  {Landauer}},\ }\href {\doibase 10.1147/rd.53.0183} {\bibfield  {journal}
  {\bibinfo  {journal} {IBM J. Res.  Dev.}\ }\textbf
  {\bibinfo {volume} {5}},\ \bibinfo {pages} {183 } (\bibinfo {year}
  {1961})}\BibitemShut {NoStop}%
\bibitem [{\citenamefont {Caves}(1990)}]{caves}%
  \BibitemOpen
  \bibfield  {author} {\bibinfo {author} {\bibfnamefont {C.~M.}\ \bibnamefont
  {Caves}},\ }\href {\doibase 10.1103/PhysRevLett.64.2111} {\bibfield
  {journal} {\bibinfo  {journal} {Phys. Rev. Lett.}\ }\textbf {\bibinfo
  {volume} {64}},\ \bibinfo {pages} {2111} (\bibinfo {year}
  {1990})}\BibitemShut {NoStop}%
\bibitem [{\citenamefont {Scully}(2001)}]{scullyPRL2001}%
  \BibitemOpen
  \bibfield  {author} {\bibinfo {author} {\bibfnamefont {M.~O.}\ \bibnamefont
  {Scully}},\ }\href {\doibase 10.1103/PhysRevLett.87.220601} {\bibfield
  {journal} {\bibinfo  {journal} {Phys. Rev. Lett.}\ }\textbf {\bibinfo
  {volume} {87}},\ \bibinfo {pages} {220601} (\bibinfo {year}
  {2001})}\BibitemShut {NoStop}%
\bibitem [{\citenamefont {Toyabe}\ \emph {et~al.}(2010)\citenamefont {Toyabe},
  \citenamefont {Sagawa}, \citenamefont {Ueda}, \citenamefont {Muneyuki},\ and\
  \citenamefont {Sano}}]{uedanature}%
  \BibitemOpen
  \bibfield  {author} {\bibinfo {author} {\bibfnamefont {S.}~\bibnamefont
  {Toyabe}}, \bibinfo {author} {\bibfnamefont {T.}~\bibnamefont {Sagawa}},
  \bibinfo {author} {\bibfnamefont {M.}~\bibnamefont {Ueda}}, \bibinfo {author}
  {\bibfnamefont {E.}~\bibnamefont {Muneyuki}}, \ and\ \bibinfo {author}
  {\bibfnamefont {M.}~\bibnamefont {Sano}},\ }\href {\doibase
  10.1038/nphys1821} {\bibfield  {journal} {\bibinfo  {journal} {Nature
  Physics}\ }\textbf {\bibinfo {volume} {6}},\ \bibinfo {pages} {988} (\bibinfo
  {year} {2010})}\BibitemShut {NoStop}%
\bibitem [{\citenamefont {Zhou}\ and\ \citenamefont {Segal}(2010)}]{segal2010}%
  \BibitemOpen
  \bibfield  {author} {\bibinfo {author} {\bibfnamefont {Y.}~\bibnamefont
  {Zhou}}\ and\ \bibinfo {author} {\bibfnamefont {D.}~\bibnamefont {Segal}},\
  }\href {\doibase 10.1103/PhysRevE.82.011120} {\bibfield  {journal} {\bibinfo
  {journal} {Phys. Rev. E}\ }\textbf {\bibinfo {volume} {82}},\ \bibinfo
  {pages} {011120} (\bibinfo {year} {2010})}\BibitemShut {NoStop}%
\bibitem [{\citenamefont {Maruyama}\ \emph {et~al.}(2009)\citenamefont
  {Maruyama}, \citenamefont {Nori},\ and\ \citenamefont
  {Vedral}}]{MaruyamaRMP2009}%
  \BibitemOpen
  \bibfield  {author} {\bibinfo {author} {\bibfnamefont {K.}~\bibnamefont
  {Maruyama}}, \bibinfo {author} {\bibfnamefont {F.}~\bibnamefont {Nori}}, \
  and\ \bibinfo {author} {\bibfnamefont {V.}~\bibnamefont {Vedral}},\ }\href
  {\doibase 10.1103/RevModPhys.81.1} {\bibfield  {journal} {\bibinfo  {journal}
  {Rev. Mod. Phys.}\ }\textbf {\bibinfo {volume} {81}},\ \bibinfo {pages} {1}
  (\bibinfo {year} {2009})}\BibitemShut {NoStop}%
\bibitem [{\citenamefont {Ruschhaupt}\ \emph {et~al.}(2006)\citenamefont
  {Ruschhaupt}, \citenamefont {Muga},\ and\ \citenamefont
  {Raizen}}]{Ruschhauptjphisb06}%
  \BibitemOpen
  \bibfield  {author} {\bibinfo {author} {\bibfnamefont {A.}~\bibnamefont
  {Ruschhaupt}}, \bibinfo {author} {\bibfnamefont {J.~G.}\ \bibnamefont
  {Muga}}, \ and\ \bibinfo {author} {\bibfnamefont {M.~G.}\ \bibnamefont
  {Raizen}},\ }\href {http://stacks.iop.org/0953-4075/39/i=18/a=012} {\bibfield
   {journal} {\bibinfo  {journal} {Journal of Physics B}\ }\textbf {\bibinfo
  {volume} {39}},\ \bibinfo {pages} {3833} (\bibinfo {year}
  {2006})}\BibitemShut {NoStop}%
\bibitem [{\citenamefont {More}\ and\ \citenamefont {Scully}(1984)}]{morebook}%
  \BibitemOpen
  \bibinfo {editor} {\bibfnamefont {G.~T.}\ \bibnamefont {More}}\ and\ \bibinfo
  {editor} {\bibfnamefont {M.~O.}\ \bibnamefont {Scully}},\ eds.,\ \href@noop
  {} {\emph {\bibinfo {title} {Frontiers of Non-Equilibrium Statistical
  Physics}}}\ (\bibinfo  {publisher} {Plenum Press},\ \bibinfo {year}
  {1984})\BibitemShut {NoStop}%
\bibitem [{\citenamefont {Sagawa}\ and\ \citenamefont
  {Ueda}(2008)}]{uedatheory}%
  \BibitemOpen
  \bibfield  {author} {\bibinfo {author} {\bibfnamefont {T.}~\bibnamefont
  {Sagawa}}\ and\ \bibinfo {author} {\bibfnamefont {M.}~\bibnamefont {Ueda}},\
  }\href {\doibase 10.1103/PhysRevLett.100.080403} {\bibfield  {journal}
  {\bibinfo  {journal} {Phys. Rev. Lett.}\ }\textbf {\bibinfo {volume} {100}},\
  \bibinfo {pages} {080403} (\bibinfo {year} {2008})}\BibitemShut {NoStop}%
\bibitem [{\citenamefont {Jacobs}(2012)}]{JacobsPRE2012}%
  \BibitemOpen
  \bibfield  {author} {\bibinfo {author} {\bibfnamefont {K.}~\bibnamefont
  {Jacobs}},\ }\href {\doibase 10.1103/PhysRevE.86.040106} {\bibfield
  {journal} {\bibinfo  {journal} {Phys. Rev. E}\ }\textbf {\bibinfo {volume}
  {86}},\ \bibinfo {pages} {040106} (\bibinfo {year} {2012})}\BibitemShut
  {NoStop}%
\bibitem [{\citenamefont {Rio}\ \emph {et~al.}(2011)\citenamefont {Rio},
  \citenamefont {Aberg}, \citenamefont {Renner}, \citenamefont {Dahlsten},\
  and\ \citenamefont {Vedral}}]{delrioNAT11}%
  \BibitemOpen
  \bibfield  {author} {\bibinfo {author} {\bibfnamefont {L.~d.}\ \bibnamefont
  {Rio}}, \bibinfo {author} {\bibfnamefont {J.}~\bibnamefont {Aberg}}, \bibinfo
  {author} {\bibfnamefont {R.}~\bibnamefont {Renner}}, \bibinfo {author}
  {\bibfnamefont {O.}~\bibnamefont {Dahlsten}}, \ and\ \bibinfo {author}
  {\bibfnamefont {V.}~\bibnamefont {Vedral}},\ }\href@noop {} {\bibfield
  {journal} {\bibinfo  {journal} {Nature}\ }\textbf {\bibinfo {volume} {474}},\
  \bibinfo {pages} {61} (\bibinfo {year} {2011})}\BibitemShut {NoStop}%
\bibitem [{\citenamefont {Allahverdyan}\ and\ \citenamefont
  {Nieuwenhuizen}(2005)}]{AllahverdyanPRE05}%
  \BibitemOpen
  \bibfield  {author} {\bibinfo {author} {\bibfnamefont {A.~E.}\ \bibnamefont
  {Allahverdyan}}\ and\ \bibinfo {author} {\bibfnamefont {T.~M.}\ \bibnamefont
  {Nieuwenhuizen}},\ }\href@noop {} {\bibfield  {journal} {\bibinfo  {journal}
  {Phys. Rev. E}\ }\textbf {\bibinfo {volume} {71}},\ \bibinfo {pages} {046107}
  (\bibinfo {year} {2005})}\BibitemShut {NoStop}%
\bibitem [{\citenamefont {Alicki}(1979)}]{Alickiwork}%
  \BibitemOpen
  \bibfield  {author} {\bibinfo {author} {\bibfnamefont {R.}~\bibnamefont
  {Alicki}},\ }\href@noop {} {\bibfield  {journal} {\bibinfo  {journal}
  {Journal of Physics A}\ }\textbf {\bibinfo {volume} {12}},\ \bibinfo {pages}
  {L103} (\bibinfo {year} {1979})}\BibitemShut {NoStop}%
\bibitem [{\citenamefont {Lindblad}(1983)}]{Lindbladbook}%
  \BibitemOpen
  \bibfield  {author} {\bibinfo {author} {\bibfnamefont {G.}~\bibnamefont
  {Lindblad}},\ }\href@noop {} {\emph {\bibinfo {title} {Non-Equilibrium
  Entropy and Irreversibility}}}\ (\bibinfo  {publisher} {D. Reidel},\ \bibinfo
  {address} {Holland},\ \bibinfo {year} {1983})\BibitemShut {NoStop}%
\bibitem [{\citenamefont {Erez}\ \emph {et~al.}(2008)\citenamefont {Erez},
  \citenamefont {Gordon}, \citenamefont {Nest},\ and\ \citenamefont
  {Kurizki}}]{nature}%
  \BibitemOpen
  \bibfield  {author} {\bibinfo {author} {\bibfnamefont {N.}~\bibnamefont
  {Erez}}, \bibinfo {author} {\bibfnamefont {G.}~\bibnamefont {Gordon}},
  \bibinfo {author} {\bibfnamefont {M.}~\bibnamefont {Nest}}, \ and\ \bibinfo
  {author} {\bibfnamefont {G.}~\bibnamefont {Kurizki}},\ }\href {\doibase
  10.1038/nature06873} {\bibfield  {journal} {\bibinfo  {journal} {Nature}\
  }\textbf {\bibinfo {volume} {452}},\ \bibinfo {pages} {724} (\bibinfo {year}
  {2008})}\BibitemShut {NoStop}%
\bibitem [{\citenamefont {Chapin}\ and\ \citenamefont
  {Scully}(2008)}]{chapinnat08}%
  \BibitemOpen
  \bibfield  {author} {\bibinfo {author} {\bibfnamefont {R.}~\bibnamefont
  {Chapin}}\ and\ \bibinfo {author} {\bibfnamefont {M.}~\bibnamefont
  {Scully}},\ }\href@noop {} {\bibfield  {journal} {\bibinfo  {journal}
  {Nature}\ }\textbf {\bibinfo {volume} {452}},\ \bibinfo {pages} {705}
  (\bibinfo {year} {2008})}\BibitemShut {NoStop}%
\bibitem [{\citenamefont {Gordon}\ \emph {et~al.}(2009)\citenamefont {Gordon},
  \citenamefont {Bensky}, \citenamefont {Gelbwaser-Klimovsky}, \citenamefont
  {Rao}, \citenamefont {Erez},\ and\ \citenamefont {Kurizki}}]{njp}%
  \BibitemOpen
  \bibfield  {author} {\bibinfo {author} {\bibfnamefont {G.}~\bibnamefont
  {Gordon}}, \bibinfo {author} {\bibfnamefont {G.}~\bibnamefont {Bensky}},
  \bibinfo {author} {\bibfnamefont {D.}~\bibnamefont {Gelbwaser-Klimovsky}},
  \bibinfo {author} {\bibfnamefont {D.~D.~B.}\ \bibnamefont {Rao}}, \bibinfo
  {author} {\bibfnamefont {N.}~\bibnamefont {Erez}}, \ and\ \bibinfo {author}
  {\bibfnamefont {G.}~\bibnamefont {Kurizki}},\ }\href@noop {} {\bibfield
  {journal} {\bibinfo  {journal} {New Journal of Physics}\ }\textbf {\bibinfo
  {volume} {11}},\ \bibinfo {pages} {123025} (\bibinfo {year}
  {2009})}\BibitemShut {NoStop}%
\bibitem [{\citenamefont {Gordon}\ \emph {et~al.}(2010)\citenamefont {Gordon},
  \citenamefont {Rao},\ and\ \citenamefont {Kurizki}}]{njp10}%
  \BibitemOpen
  \bibfield  {author} {\bibinfo {author} {\bibfnamefont {G.}~\bibnamefont
  {Gordon}}, \bibinfo {author} {\bibfnamefont {D.~D.~B.}\ \bibnamefont {Rao}},
  \ and\ \bibinfo {author} {\bibfnamefont {G.}~\bibnamefont {Kurizki}},\
  }\href@noop {} {\bibfield  {journal} {\bibinfo  {journal} {New Journal of
  Physics}\ }\textbf {\bibinfo {volume} {12}},\ \bibinfo {pages} {053033}
  (\bibinfo {year} {2010})}\BibitemShut {NoStop}%
\bibitem [{\citenamefont {Alvarez}\ \emph {et~al.}(2010)\citenamefont
  {Alvarez}, \citenamefont {Rao}, \citenamefont {Frydman},\ and\ \citenamefont
  {Kurizki}}]{Gonzalo}%
  \BibitemOpen
  \bibfield  {author} {\bibinfo {author} {\bibfnamefont {G.~A.}\ \bibnamefont
  {Alvarez}}, \bibinfo {author} {\bibfnamefont {D.~D.~B.}\ \bibnamefont {Rao}},
  \bibinfo {author} {\bibfnamefont {L.}~\bibnamefont {Frydman}}, \ and\
  \bibinfo {author} {\bibfnamefont {G.}~\bibnamefont {Kurizki}},\ }\href
  {\doibase 10.1103/PhysRevLett.105.160401} {\bibfield  {journal} {\bibinfo
  {journal} {Phys. Rev. Lett.}\ }\textbf {\bibinfo {volume} {105}},\ \bibinfo
  {pages} {160401} (\bibinfo {year} {2010})}\BibitemShut {NoStop}%
\bibitem [{\citenamefont {Jahnke}\ and\ \citenamefont {Mahler}(2010)}]{Jahnke}%
  \BibitemOpen
  \bibfield  {author} {\bibinfo {author} {\bibfnamefont {T.}~\bibnamefont
  {Jahnke}}\ and\ \bibinfo {author} {\bibfnamefont {G.}~\bibnamefont
  {Mahler}},\ }\href@noop {} {\bibfield  {journal} {\bibinfo  {journal} {EPL
  (Europhysics Letters)}\ }\textbf {\bibinfo {volume} {90}},\ \bibinfo {pages}
  {50008} (\bibinfo {year} {2010})}\BibitemShut {NoStop}%
\bibitem [{\citenamefont {Scully}\ \emph {et~al.}(2011)\citenamefont {Scully},
  \citenamefont {Chapin}, \citenamefont {Dorfman}, \citenamefont {Kim},\ and\
  \citenamefont {Svidzinsky}}]{Scully2011}%
  \BibitemOpen
  \bibfield  {author} {\bibinfo {author} {\bibfnamefont {M.~O.}\ \bibnamefont
  {Scully}}, \bibinfo {author} {\bibfnamefont {K.~R.}\ \bibnamefont {Chapin}},
  \bibinfo {author} {\bibfnamefont {K.~E.}\ \bibnamefont {Dorfman}}, \bibinfo
  {author} {\bibfnamefont {M.~B.}\ \bibnamefont {Kim}}, \ and\ \bibinfo
  {author} {\bibfnamefont {A.}~\bibnamefont {Svidzinsky}},\ }\href {\doibase
  10.1073/pnas.1110234108} {\bibfield  {journal} {\bibinfo  {journal} {PNAS}\
  }\textbf {\bibinfo {volume} {108}},\ \bibinfo {pages} {15097} (\bibinfo
  {year} {2011})}\BibitemShut {NoStop}%
\bibitem [{\citenamefont {Scully}\ \emph {et~al.}(2003)\citenamefont {Scully},
  \citenamefont {Zubairy}, \citenamefont {Agarwal},\ and\ \citenamefont
  {Walther}}]{Scullysinglebath}%
  \BibitemOpen
  \bibfield  {author} {\bibinfo {author} {\bibfnamefont {M.~O.}\ \bibnamefont
  {Scully}}, \bibinfo {author} {\bibfnamefont {M.~S.}\ \bibnamefont {Zubairy}},
  \bibinfo {author} {\bibfnamefont {G.~S.}\ \bibnamefont {Agarwal}}, \ and\
  \bibinfo {author} {\bibfnamefont {H.}~\bibnamefont {Walther}},\ }\href
  {\doibase 10.1126/science.1078955} {\bibfield  {journal} {\bibinfo  {journal}
  {Science}\ }\textbf {\bibinfo {volume} {299}},\ \bibinfo {pages} {862}
  (\bibinfo {year} {2003})}\BibitemShut {NoStop}%
\bibitem [{\citenamefont {Scully}(2010)}]{scullysinglebath2}%
  \BibitemOpen
  \bibfield  {author} {\bibinfo {author} {\bibfnamefont {M.~O.}\ \bibnamefont
  {Scully}},\ }\href {\doibase 10.1103/PhysRevLett.104.207701} {\bibfield
  {journal} {\bibinfo  {journal} {Phys. Rev. Lett.}\ }\textbf {\bibinfo
  {volume} {104}},\ \bibinfo {pages} {207701} (\bibinfo {year}
  {2010})}\BibitemShut {NoStop}%
\bibitem [{\citenamefont {Scully}\ and\ \citenamefont
  {Zubairy}(1997)}]{ScullyBOOK97}%
  \BibitemOpen
  \bibfield  {author} {\bibinfo {author} {\bibfnamefont {M.~O.}\ \bibnamefont
  {Scully}}\ and\ \bibinfo {author} {\bibfnamefont {M.~S.}\ \bibnamefont
  {Zubairy}},\ }\href@noop {} {\emph {\bibinfo {title} {Quantum Optics}}}\
  (\bibinfo  {publisher} {Cambridge University Press},\ \bibinfo {address}
  {Cambridge, UK},\ \bibinfo {year} {1997})\BibitemShut {NoStop}%
\bibitem [{\citenamefont {Alicki}\ \emph {et~al.}(2004)\citenamefont {Alicki},
  \citenamefont {Horodecki}, \citenamefont {Horodecki},\ and\ \citenamefont
  {Horodecki}}]{alicki2004thermodynamics}%
  \BibitemOpen
  \bibfield  {author} {\bibinfo {author} {\bibfnamefont {R.}~\bibnamefont
  {Alicki}}, \bibinfo {author} {\bibfnamefont {M.}~\bibnamefont {Horodecki}},
  \bibinfo {author} {\bibfnamefont {P.}~\bibnamefont {Horodecki}}, \ and\
  \bibinfo {author} {\bibfnamefont {R.}~\bibnamefont {Horodecki}},\ }\href@noop
  {} {\bibfield  {journal} {\bibinfo  {journal} {Open Systems \& Information
  Dynamics}\ }\textbf {\bibinfo {volume} {11}},\ \bibinfo {pages} {205}
  (\bibinfo {year} {2004})}\BibitemShut {NoStop}%
\bibitem [{\citenamefont {Breuer}\ and\ \citenamefont
  {Petruccione}(2002)}]{petruccione}%
  \BibitemOpen
  \bibfield  {author} {\bibinfo {author} {\bibfnamefont {H.-P.}\ \bibnamefont
  {Breuer}}\ and\ \bibinfo {author} {\bibfnamefont {F.}~\bibnamefont
  {Petruccione}},\ }\href@noop {} {\emph {\bibinfo {title} {The theory of open
  quantum systems}}}\ (\bibinfo  {publisher} {Oxford},\ \bibinfo {year}
  {2002})\BibitemShut {NoStop}%
\bibitem [{\citenamefont {Kofman}\ \emph {et~al.}(1994)\citenamefont {Kofman},
  \citenamefont {Kurizki},\ and\ \citenamefont {Sherman}}]{dressedstates}%
  \BibitemOpen
  \bibfield  {author} {\bibinfo {author} {\bibfnamefont {A.~G.}\ \bibnamefont
  {Kofman}}, \bibinfo {author} {\bibfnamefont {G.}~\bibnamefont {Kurizki}}, \
  and\ \bibinfo {author} {\bibfnamefont {B.}~\bibnamefont {Sherman}},\
  }\href@noop {} {\bibfield  {journal} {\bibinfo  {journal} {Journal of Modern
  Optics}\ }\textbf {\bibinfo {volume} {41}},\ \bibinfo {pages} {353} (\bibinfo
  {year} {1994})}\BibitemShut {NoStop}%
\bibitem [{\citenamefont {Kofman}\ and\ \citenamefont
  {Kurizki}(1996)}]{kofPRA96}%
  \BibitemOpen
  \bibfield  {author} {\bibinfo {author} {\bibfnamefont {A.~G.}\ \bibnamefont
  {Kofman}}\ and\ \bibinfo {author} {\bibfnamefont {G.}~\bibnamefont
  {Kurizki}},\ }\href {\doibase 10.1103/PhysRevA.54.R3750} {\bibfield
  {journal} {\bibinfo  {journal} {Phys. Rev. A}\ }\textbf {\bibinfo {volume}
  {54}},\ \bibinfo {pages} {R3750} (\bibinfo {year} {1996})}\BibitemShut
  {NoStop}%
\bibitem [{\citenamefont {Kofman}\ and\ \citenamefont
  {Kurizki}(2004)}]{kofmantemp}%
  \BibitemOpen
  \bibfield  {author} {\bibinfo {author} {\bibfnamefont {A.~G.}\ \bibnamefont
  {Kofman}}\ and\ \bibinfo {author} {\bibfnamefont {G.}~\bibnamefont
  {Kurizki}},\ }\href {\doibase 10.1103/PhysRevLett.93.130406} {\bibfield
  {journal} {\bibinfo  {journal} {Phys. Rev. Lett.}\ }\textbf {\bibinfo
  {volume} {93}},\ \bibinfo {pages} {130406} (\bibinfo {year}
  {2004})}\BibitemShut {NoStop}%
\bibitem [{\citenamefont {Shahmoon}\ and\ \citenamefont
  {Kurizki}(2013)}]{shahmoonPRA13}%
  \BibitemOpen
  \bibfield  {author} {\bibinfo {author} {\bibfnamefont {E.}~\bibnamefont
  {Shahmoon}}\ and\ \bibinfo {author} {\bibfnamefont {G.}~\bibnamefont
  {Kurizki}},\ }\href {\doibase 10.1103/PhysRevA.87.013841} {\bibfield
  {journal} {\bibinfo  {journal} {Phys. Rev. A}\ }\textbf {\bibinfo {volume}
  {87}},\ \bibinfo {pages} {013841} (\bibinfo {year} {2013})}\BibitemShut
  {NoStop}%
\bibitem [{\citenamefont {Nieuwenhuizen}\ and\ \citenamefont
  {Allahverdyan}(2002)}]{niewuallah}%
  \BibitemOpen
  \bibfield  {author} {\bibinfo {author} {\bibfnamefont {T.~M.}\ \bibnamefont
  {Nieuwenhuizen}}\ and\ \bibinfo {author} {\bibfnamefont {A.~E.}\ \bibnamefont
  {Allahverdyan}},\ }\href {\doibase 10.1103/PhysRevE.66.036102} {\bibfield
  {journal} {\bibinfo  {journal} {Phys. Rev. E}\ }\textbf {\bibinfo {volume}
  {66}},\ \bibinfo {pages} {036102} (\bibinfo {year} {2002})}\BibitemShut
  {NoStop}%
\bibitem [{\citenamefont {Petrosyan}\ \emph {et~al.}(2009)\citenamefont
  {Petrosyan}, \citenamefont {Bensky}, \citenamefont {Kurizki}, \citenamefont
  {Mazets}, \citenamefont {Majer},\ and\ \citenamefont
  {Schmiedmayer}}]{PetrosyanPRA09}%
  \BibitemOpen
  \bibfield  {author} {\bibinfo {author} {\bibfnamefont {D.}~\bibnamefont
  {Petrosyan}}, \bibinfo {author} {\bibfnamefont {G.}~\bibnamefont {Bensky}},
  \bibinfo {author} {\bibfnamefont {G.}~\bibnamefont {Kurizki}}, \bibinfo
  {author} {\bibfnamefont {I.}~\bibnamefont {Mazets}}, \bibinfo {author}
  {\bibfnamefont {J.}~\bibnamefont {Majer}}, \ and\ \bibinfo {author}
  {\bibfnamefont {J.}~\bibnamefont {Schmiedmayer}},\ }\href {\doibase
  10.1103/PhysRevA.79.040304} {\bibfield  {journal} {\bibinfo  {journal} {Phys.
  Rev. A}\ }\textbf {\bibinfo {volume} {79}},\ \bibinfo {pages} {040304}
  (\bibinfo {year} {2009})}\BibitemShut {NoStop}%
\bibitem [{\citenamefont {Verd\'u}\ \emph {et~al.}(2009)\citenamefont
  {Verd\'u}, \citenamefont {Zoubi}, \citenamefont {Koller}, \citenamefont
  {Majer}, \citenamefont {Ritsch},\ and\ \citenamefont
  {Schmiedmayer}}]{nvcenter}%
  \BibitemOpen
  \bibfield  {author} {\bibinfo {author} {\bibfnamefont {J.}~\bibnamefont
  {Verd\'u}}, \bibinfo {author} {\bibfnamefont {H.}~\bibnamefont {Zoubi}},
  \bibinfo {author} {\bibfnamefont {C.}~\bibnamefont {Koller}}, \bibinfo
  {author} {\bibfnamefont {J.}~\bibnamefont {Majer}}, \bibinfo {author}
  {\bibfnamefont {H.}~\bibnamefont {Ritsch}}, \ and\ \bibinfo {author}
  {\bibfnamefont {J.}~\bibnamefont {Schmiedmayer}},\ }\href {\doibase
  10.1103/PhysRevLett.103.043603} {\bibfield  {journal} {\bibinfo  {journal}
  {Phys. Rev. Lett.}\ }\textbf {\bibinfo {volume} {103}},\ \bibinfo {pages}
  {043603} (\bibinfo {year} {2009})}\BibitemShut {NoStop}%
\bibitem [{\citenamefont {Scovil}\ and\ \citenamefont
  {Schulz-DuBois}(1959)}]{ScovilPRL59}%
  \BibitemOpen
  \bibfield  {author} {\bibinfo {author} {\bibfnamefont {H.~E.~D.}\
  \bibnamefont {Scovil}}\ and\ \bibinfo {author} {\bibfnamefont {E.~O.}\
  \bibnamefont {Schulz-DuBois}},\ }\href {\doibase 10.1103/PhysRevLett.2.262}
  {\bibfield  {journal} {\bibinfo  {journal} {Phys. Rev. Lett.}\ }\textbf
  {\bibinfo {volume} {2}},\ \bibinfo {pages} {262} (\bibinfo {year}
  {1959})}\BibitemShut {NoStop}%
\bibitem [{\citenamefont {Ford}\ and\ \citenamefont
  {O'Connell}(2006)}]{FordPRL06}%
  \BibitemOpen
  \bibfield  {author} {\bibinfo {author} {\bibfnamefont {G.~W.}\ \bibnamefont
  {Ford}}\ and\ \bibinfo {author} {\bibfnamefont {R.~F.}\ \bibnamefont
  {O'Connell}},\ }\href {\doibase 10.1103/PhysRevLett.96.020402} {\bibfield
  {journal} {\bibinfo  {journal} {Phys. Rev. Lett.}\ }\textbf {\bibinfo
  {volume} {96}},\ \bibinfo {pages} {020402} (\bibinfo {year}
  {2006})}\BibitemShut {NoStop}%
\bibitem [{\citenamefont {Bhaktavatsala~Rao}\ and\ \citenamefont
  {Kurizki}(2011)}]{durgaphysreva}%
  \BibitemOpen
  \bibfield  {author} {\bibinfo {author} {\bibfnamefont {D.~D.}\ \bibnamefont
  {Bhaktavatsala~Rao}}\ and\ \bibinfo {author} {\bibfnamefont {G.}~\bibnamefont
  {Kurizki}},\ }\href {\doibase 10.1103/PhysRevA.83.032105} {\bibfield
  {journal} {\bibinfo  {journal} {Phys. Rev. A}\ }\textbf {\bibinfo {volume}
  {83}},\ \bibinfo {pages} {032105} (\bibinfo {year} {2011})}\BibitemShut
  {NoStop}%
\bibitem [{\citenamefont {Erez}(2012)}]{maxwork}%
  \BibitemOpen
  \bibfield  {author} {\bibinfo {author} {\bibfnamefont {N.}~\bibnamefont
  {Erez}},\ }\href {http://stacks.iop.org/1402-4896/2012/i=T151/a=014028}
  {\bibfield  {journal} {\bibinfo  {journal} {Physica Scripta}\ }\textbf
  {\bibinfo {volume} {2012}},\ \bibinfo {pages} {014028} (\bibinfo {year}
  {2012})}\BibitemShut {NoStop}%
\bibitem [{\citenamefont {Takara}\ \emph {et~al.}(2010)\citenamefont {Takara},
  \citenamefont {Hasegawa},\ and\ \citenamefont {Driebe}}]{Takara201088}%
  \BibitemOpen
  \bibfield  {author} {\bibinfo {author} {\bibfnamefont {K.}~\bibnamefont
  {Takara}}, \bibinfo {author} {\bibfnamefont {H.-H.}\ \bibnamefont
  {Hasegawa}}, \ and\ \bibinfo {author} {\bibfnamefont {D.}~\bibnamefont
  {Driebe}},\ }\href {\doibase DOI: 10.1016/j.physleta.2010.11.002} {\bibfield
  {journal} {\bibinfo  {journal} {Physics Letters A}\ }\textbf {\bibinfo
  {volume} {375}},\ \bibinfo {pages} {88 } (\bibinfo {year}
  {2010})}\BibitemShut {NoStop}%
\end{thebibliography}
\end{document}